\documentclass[draft]{agujournal2019}
\usepackage{url} 
\usepackage{lineno}
\usepackage{booktabs}
\usepackage{amsmath}
\usepackage[inline]{trackchanges} 
\usepackage{soul}
\draftfalse

\journalname{Journal of Advances in Modeling Earth Systems (JAMES)}

\begin{document}

%
%


\title{climt-paraformer: Stable Emulation of Convective Parameterization using a Temporal Memory-aware Transformer}

%
%




\authors{Shuochen Wang\affil{1}, Nishant Yadav\affil{2}, Joy Merwin Monteiro\affil{3}, Auroop R. Ganguly\affil{1,4}}


\affiliation{1}{Sustainability and Data Sciences Laboratory, Northeastern University}
\affiliation{2}{Microsoft}
\affiliation{3}{Indian Institute of Science Education and Research Pune}
\affiliation{4}{AI for Climate and Sustainability, The Institute for Experiential AI, Northeastern University}




\correspondingauthor{Auroop R. Ganguly}{a.ganguly@northeastern.edu}



\begin{keypoints}
\item We developed a temporal memory-aware Transformer for the Emanuel convective parameterization and achieved a 10-year online simulation.
\item Modeling temporal memory with a Transformer significantly improves the representation of moist convection compared to other neural networks.
\item The sensitivity analysis shows that including about 100 minutes yields the best online performance, while longer memory degrades accuracy.
\end{keypoints}

%
%

%
%


\begin{abstract}
Accurate representation of moist convective sub-grid-scale processes remains a major challenge in global climate models, as traditional parameterization schemes are both computationally expensive and difficult to scale. Neural network (NN) emulators offer a promising alternative by learning efficient mappings between atmospheric states and convective tendencies while retaining fidelity to the underlying physics. However, most existing NN-based parameterizations are memory-less and rely only on instantaneous inputs, even though convection evolves over time and depends on prior atmospheric states. Recent studies have begun to incorporate convective memory, but they often treat past states as independent features rather than modeling temporal dependencies explicitly. In this work, we develop a temporal memory-aware Transformer emulator for the Emanuel convective parameterization and evaluate it in a single-column climate model (SCM) under both offline and online configurations. The Transformer captures temporal correlations and nonlinear interactions across consecutive atmospheric states. Compared with baseline emulators, including a memory-less multilayer perceptron and a recurrent long short-term memory model, the Transformer achieves lower offline errors. Sensitivity analysis indicates that a memory length of approximately 100 minutes yields the best performance, whereas longer memory degrades performance. We further test the emulator in long-term coupled simulations and show that it remains stable over 10 years. Among the predicted variables, moistening tendencies are more difficult to emulate than heating tendencies, and convective precipitation shows the largest discrepancy between offline and online evaluations, highlighting the difficulty of capturing regime-dependent behavior in coupled simulations. Overall, this study demonstrates the importance of explicit temporal modeling for NN-based parameterizations.
\end{abstract}

\section*{Plain Language Summary}
Climate models struggle to represent small-scale processes like moist convection, which are important for weather and climate but too complex to simulate directly. Traditional methods to approximate these processes are slow and hard to scale. Neural networks offer a faster alternative by learning how atmospheric conditions relate to heating, moisture, and rainfall. However, most existing models only look at the current state of the atmosphere, even though convection depends on what happened in the recent past. In this study, we develop a Transformer-based model that can “remember” previous atmospheric states and use that information to make better predictions. We test it in a climate model and find that it performs better than standard approaches. We also find that using about 100 minutes of past information gives the best results—too little or too much memory reduces performance. When we run long simulations, our model remains stable over 10 years, which is an important requirement for climate applications. Overall, predicting moisture changes is harder than predicting temperature changes, and rainfall is the most difficult variable to simulate accurately. This work highlights that including time history and choosing the right amount of memory are crucial for building reliable AI-based climate models.

%
%

%


%
%
%
%

\section{Introduction}
Accurately representing sub-grid-scale (SGS) physical processes remains one of the main challenges in climate and weather modeling. The current generation of Global Climate Models (GCMs) typically operates at horizontal resolutions on the order of tens to hundreds of kilometers, which is too coarse to explicitly resolve many important atmospheric processes, including moist convection \cite{arakawa1974interaction,tiedtke1989comprehensive,emanuel1991scheme}, cloud formation \cite{arakawa2004cumulus,gettelman2008new} and radiative transfer \cite{mlawer1997radiative,lacis1974parameterization}. These processes occur at spatial scales much smaller than the model's grid spacing but significantly influence the large-scale circulation. Therefore, GCMs rely on parameterization schemes that approximate the aggregate effects of SGS dynamics on resolved atmospheric variables. Among these processes, moist convection is particularly important due to its role in vertical energy transport, driving precipitation, and interacting strongly with clouds and radiation \cite{arakawa2004cumulus,emanuel1999development}. Inaccuracies in convective parameterization can introduce systematic biases in precipitation patterns, tropical variability, and large-scale circulation, ultimately affecting projections of climate sensitivity and extreme weather \cite{stevens2013climate,schneider2017climate,zelinka2020causes,cronin2017clouds}. Therefore, improving the representation of SGS physics—especially moist convection—remains essential to improve both the accuracy and numerical stability of climate simulations and weather forecasts.

Over the years, a variety of convective parameterization schemes have been developed, including convective adjustment methods and moisture-convergence schemes \cite{manabe1965simulated,kuo1974further} and more physically based frameworks such as mass-flux convection schemes \cite{arakawa1974interaction,tiedtke1989comprehensive,emanuel1991scheme,gregory1990mass,zhang1995sensitivity}. One important limitation is the assumption of quasi-equilibrium, in which convective tendencies are determined primarily by the instantaneous large-scale atmospheric state. In reality, convective processes often exhibit memory and temporal persistence, as present convective behavior depends not only on the instantaneous state but also on prior convective activity and the earlier evolution of low-level thermodynamic structure, including moisture buildup and convective inhibition \cite{colin2019identifying,tuckman2023evolution}. Neglecting these temporal dependencies can limit the ability of parameterization schemes to represent the timing and intensity of convection accurately. Recent studies have shown that incorporating convective memory, through prognostic variables or time-lagged atmospheric states, can improve the representation of convective variability and precipitation processes \cite{mapes2006mesoscale,davies2009simple}. In addition, current physics-based parameterization schemes are often computationally demanding, involving iterative calculations that must be evaluated at every model time step and grid column. As climate models move toward higher spatial resolution and longer simulations, the computational cost associated with physical parameterizations can become a significant bottleneck \cite{schneider2017climate}.

In recent years, machine learning (ML)–based emulators have emerged as a promising approach to address these challenges by learning a fast statistical approximation of existing physical parameterization schemes. Studies have shown that neural networks (NNs) can approximate the mapping between atmospheric state variables and the resulting physical tendencies, potentially reducing computational cost while maintaining comparable accuracy and stability. These experiments were conducted within an entirely idealized setting such as Lorenz 96 \cite{arnold2013stochastic,gagne2020machine}, a Cloud Resolving Model (CRM) \cite{rasp2018deep,gentine2018could,brenowitz2018prognostic,o2018using}, or other parameterization schemes used in weather and climate models \cite{krasnopolsky2013using,zhong2024machine,perkins2024emulation}. More recent studies focus on resolution generalizability \cite{yuval2020stable}, physics-informed emulators \cite{kashinath2021physics}, and achieved long-term stable online simulations 
\cite{han2025decadal,wang2026condensnet,balogh2025online}. 

Most existing NN emulators for convective parameterization rely on architectures such as Random Forest (RF) \cite{o2018using,yuval2020stable}, Multi-Layer Perceptron (MLP) \cite{gentine2018could,song2021improved,krasnopolsky2013using}, Convolutional Neural Network (CNN) \cite{bolton2019applications,larraondo2019data,hu2025stable}, and Generative Adversarial Network (GAN) \cite{gagne2020machine,nadiga2022stochastic,perezhogin2023generative} that learn a direct mapping between the instantaneous large-scale atmospheric state and the resulting convective tendencies. These architectures are attractive because they are relatively simple, computationally efficient, and easy to integrate into existing GCMs. However, such memory-less models assume that convection depends only on the current atmospheric profile, which may limit their ability to capture important temporal processes associated with convective development and organization. Ignoring the temporal dependencies can lead to errors in the timing, intensity, and variability of convective tendencies predicted by ML emulators. To address this limitation, recent work has begun incorporating sequences of atmospheric states into the model and learn dependencies across multiple time steps \cite{hu2025stable,han2020moist,han2023ensemble,behrens2025simulating}. For example, \citeA{han2020moist} trained a ResNet on a CRM and found that atmospheric states up to 1.5 hours can affect convective processes. \citeA{hu2025stable} included the convective memory, which serves as an average value of the CRM domain in a UNet architecture. \citeA{lin2025navigating} found that removing convective memory caused a substantial increase in the ensemble-average offline error. \citeA{behrens2025simulating} found that including precipitation from the previous states can improve the prediction accuracy. \citeA{shamekh2023implicit} found that integrating a simple memory process from the previous time steps in the NN can accurately capture the temporal correlation and SGS variability in convection.

In the studies discussed above, temporal information is typically incorporated by concatenating multiple previous atmospheric states as additional input features. While this approach provides the model with limited information about recent atmospheric evolution, it treats each time step as an independent input and does not explicitly represent the temporal relationships among them (which we refer to as memory in this work). Consequently, the network must implicitly infer the ordering and interactions of past states through static weights, which becomes increasingly inefficient as the temporal window expands. Moreover, concatenating multiple time steps substantially enlarges the input space, making training more challenging. In contrast, temporal neural network architectures, such as Recurrent Neural Networks (RNNs), long short-term memory networks (LSTMs) \cite{hochreiter1997long}, and Transformers \cite{vaswani2017attention}, are specifically designed to learn sequential dependencies in data. These models either maintain internal representations of past states or use attention mechanisms to identify the most relevant temporal information for prediction. As such, they offer a more natural and efficient framework for representing convective memory, with the potential to improve both the accuracy and stability of ML-based emulators in offline training and online simulation. It is worth noting that, although some sequential architectures have been explored in prior NN-based parameterization studies \cite{hafner2025representing,yao2023physics}, they have primarily been applied along the vertical dimension of a single atmospheric state to capture inter-level dependencies, rather than temporal evolution. A notable exception is \citeA{song2025physically}, where a single previous time step is incorporated into an LSTM framework; however, the explicit modeling of longer temporal sequences in convective parameterization remains largely unexplored.

In this work, we aim to explore the role of convective memory in NN–based parameterization. We proposed a temporal memory-aware Transformer emulator for the Emanuel convective parameterization scheme and performed the offline training and online simulation against other baseline models in an SCM.

\section{Data Generation}

\subsection{climt}
We evaluate our NN emulators within climt, an open-source Python toolkit designed to support the construction of flexible model hierarchies for atmospheric and climate research \cite{monteiro2018sympl}. climt is built on the Sympl modeling framework \cite{monteiro2018sympl}, which represents a climate model as a collection of components that interact through a shared atmospheric state. A key advantage of this framework is its modular design, which enables physical parameterizations, dynamical cores, and diagnostic modules to be easily interchanged and consistently coupled, even when they differ in units, variable names, or data structures. This flexibility makes climt particularly well-suited for experimenting with different parameterization schemes while keeping the core physical model as a constant module, as individual components can be isolated, compared, or replaced with minimal modifications to the overall model configuration \cite{liu2020radnet}. Consequently, climt provides a convenient and robust platform for developing and evaluating new NN emulators of atmospheric physics.

\subsection{The Emanuel Convection Scheme}
The emulation target in this study is the Emanuel convection scheme \cite{emanuel1991scheme,emanuel1999development}, which is a widely used parameterization designed to represent the effects of deep moist convection on the large-scale atmosphere in weather and climate models such as the Regional Climate Model version 4 (RegCM4) \cite{elguindi2014regional}. The scheme is based on a mass-flux framework that represents convective clouds as ensembles of buoyant updrafts and compensating downdrafts that transport heat, moisture, and momentum vertically within the atmospheric column. It explicitly accounts for processes such as environmental air entrainment and detrainment, convective triggering based on atmospheric instability, and the interaction between convective plumes and the surrounding environment. By computing the resulting tendencies associated with convective mixing and precipitation formation, the Emanuel scheme provides a physical representation of how unresolved convective processes influence large-scale atmospheric thermodynamic profiles.

\subsection{Climate Model Configuration}
We choose a single-column radiative–convective model with a slab ocean surface in climt. A slab ocean mixed layer of 5 m depth is used to allow relatively fast convergence to radiative–convective equilibrium (RCE). The model is integrated with a 10-minute time step using the Adams–Bashforth time-stepping scheme, with the primary physical components consisting of the Rapid Radiative Transfer Model for GCMs (RRTMG) \cite{iacono2008radiative} handling the longwave and shortwave radiative parameterization and the Emanuel convection scheme \cite{emanuel1999development} for deep moist convection. After computing the coupled radiative and convective tendencies, we apply an additional bulk-physics component to represent near-surface and lower-tropospheric processes that are not explicitly handled by the main parameterizations. This physics package calculates surface sensible and latent heat fluxes, includes a simple planetary boundary layer representation, and applies heating and moistening tendencies associated with large-scale condensation under saturation conditions. A detailed description of this configuration can be found in \citeA{reed2012idealized}. The model also imposes a fixed wind speed of 3 m/s at every time step and uses pressure as the vertical coordinate, with 28 vertical levels spanning from 1013.2 hPa at the surface to 0 hPa at the model top.

The model reaches RCE after approximately four years of integration. The simulation is then continued for a total of 20 years, yielding 1,030,384 atmospheric column samples.

\section{Neural Network Setup}
In this study, we implemented 3 types of NN architectures: MLP, LSTM and Transformer. For MLP, the number of hidden layers and neurons is searched dynamically. After each hidden layer, the Rectified Linear Unit (ReLU) activation function is applied. MLP only takes one atmospheric state as input, and it is arguably the most commonly used architecture in NN parameterization. However, it is inherently memory-less and cannot explicitly represent temporal dependencies. The LSTM uses a recurrent architecture that maintains an internal memory state through gated mechanisms, allowing information to persist across multiple time steps and making it well-suited for modeling sequential atmospheric processes \cite{hochreiter1997long}. However, LSTM models are known to be less effective for long-term sequence modeling, as they can suffer from vanishing gradients that limit their ability to capture long-range dependencies \cite{al2023lstm}. The Transformer, originally developed for natural language processing \cite{vaswani2017attention}, represents a newer class of sequence models that rely on self-attention mechanisms to capture dependencies across time without recurrence. Due to its ability to model long-range temporal relationships, the Transformer has recently been applied to time-series prediction problems, including geophysical applications \cite{zhou2021informer, zerveas2021transformer, wen2022transformers}.

Both the LSTM and Transformer explicitly incorporate convective memory by taking multiple consecutive atmospheric states as input. We denote $T_w$ as the number of input states, such that the total memory length corresponds to $T_w$ multiplied by the model time step (10 minutes in this study). In contrast, the MLP corresponds to the special case $T_w$ = 1. In Transformer, $T_w$ is the sequence length, which is the number of tokens in an input sequence that the model processes simultaneously. It defines the context window—how many previous atmospheric states the model can "see" at once. The input is first linearly projected into a higher-dimensional latent space, a process called input embedding, and then augmented with sinusoidal positional encoding to represent the temporal order of the input sequence. A stack of Transformer encoder layers then models temporal dependencies using multi-head self-attention. In this context, the attention mechanism allows the model to dynamically weigh the importance of different time steps in the input sequence when making predictions, enabling it to capture both short- and long-range dependencies. By comparing each time step with all others, the model can identify relevant patterns and interactions across the sequence. The final prediction is generated by selecting the last time-step representation and passing it through a linear decoder to produce. Given that the input consists of a short sequence of atmospheric states, we also apply a causal mask to ensure that predictions at a given time step depend only on current and past inputs, thereby preserving temporal consistency. The encoder uses the Gaussian Error Linear Unit (GeLU) activation function and adopts a pre-layer normalization configuration, where layer normalization is applied before the attention and feedforward layers to improve training stability.

\section{Training}
We train the emulators using data sampled after the system has reached RCE. From the full set of generated atmospheric columns, we randomly select 400,000 samples for training, as we find that including additional samples does not improve offline performance—likely because the data are collected from a stationary RCE distribution—while substantially increasing computational cost. The input and output variables are listed in Table \ref{tab:1}. Air pressure is excluded from the inputs, as it remains constant in time under the single-column configuration. The selected samples are divided into training (40\%), validation (20\%), and test (20\%) subsets.

\begin{table}[htbp]
\centering
\caption{Input ($X$) and output ($Y$) variables used for training the NN emulators. Variables include both vertically resolved atmospheric profiles and scalar surface quantities. All variables are concatenated along the vertical dimension to construct the input and output feature vectors.}
\label{tab:1}
\resizebox{\textwidth}{!}{%
\begin{tabular}{@{}llll@{}}
\toprule
\textbf{Input ($X$)} & \textbf{Size} & \textbf{Output ($Y$)}     & \textbf{Size} \\ \midrule
Temperature, $T$ {[}K{]}                   & 28 & Heating tendency, $dT/dt$ {[}K/s{]}                & 28 \\
Specific humidity, $Q$ {[}kg/kg{]}         & 28   & Moistening tendency, $dq/dt$ {[}kg/kg/s{]} & 28   \\

Surface latent heat flux, $LHF$ {[}W/m\textsuperscript{2}{]}   & 1  & Convective precipitation rate, $precip$ {[}mm/day{]}                       & 1  \\
Surface sensible heat flux, $SHF$ {[}W/m\textsuperscript{2}{]}   & 1  &   &   \\
Cloud-base mass flux, $CLD$ {[kg/m\textsuperscript{2}/s]}   & 1  &   &   \\
\midrule                                      
\textbf{Total}                                  & 59  & \textbf{Total}                                              & 57  \\
\bottomrule
\end{tabular}%
}
\end{table}

\subsection{Normalization}
Input variables are normalized using min–max scaling, while output variables are standardized using the mean–standard-deviation method, as shown in Equation \ref{eq:norm}. In the context of moist convection parameterization, certain variables—such as specific humidity and its tendencies—can be exactly zero when convection is not triggered at a given time step or vertical level. These zero values are preserved and excluded from the normalization procedure, so that only nonzero values are scaled. All normalization statistics are computed independently at each vertical level $k$ from the training set.
\begin{equation}
X_k = \frac{X_k - X_{\min,k}}{X_{\max,k} - X_{\min,k}}, \quad
Y_k = \frac{Y_k - \bar{Y}_k}{{\sigma}_k}
\label{eq:norm}
\end{equation}

\subsection{Physical Constraints}
To improve physical consistency and reduce errors in the emulated convective tendencies, we impose hard vertical constraints on the NN inputs and outputs. Let the input temperature and specific humidity profiles of a total levels of $N = 28$ be written as $T = (T_0, T_1, ..., T_{N-1})$ and $Q = (Q_0, Q_1, ..., Q_{N-1})$, and let the predicted convective tendencies be $dT/dt$ and $dQ/dt$. As in previous studies \cite{song2025physically, hu2025stable}, we define a cutoff level $k_0$, above which convective effects are forcibly suppressed. Thermodynamic inputs are set to zero for all levels above this cutoff, and for outputs, the predicted convective tendencies are also set to zero above the same level:
\begin{equation}
T_k = 0, \quad \frac{dT_k}{dt} = 0, \quad Q_k = 0, \quad \frac{dQ_k}{dt} = 0 \quad \text{for } k \ge k_0
\label{eq:tq}
\end{equation}

These constraints are applied directly during each forward pass, so they act as strict structural priors rather than learned penalties.

In this study, the MLP is tested in both an unconstrained version ($k_0 = N$) and a constrained version, whereas the LSTM and Transformer are implemented only in the constrained form. This choice is motivated by preliminary experiments showing that the unconstrained MLP produces substantially larger errors, which will be discussed further in the offline result section. One likely reason is that target tendencies in the upper atmosphere are extremely small, so normalization can amplify numerical artifacts and make these weak signals difficult for the network to learn. By enforcing zero tendencies above $k_0$, the constrained models avoid fitting spurious variations in levels where convection should be negligible. Importantly, we do not prescribe an arbitrary cutoff level or restrict the training data to a part of layers such as in \citeA{bolton2019applications,hu2025stable,song2025physically}, since the vertical extent of convective activity depends on the specific parameterization and environmental conditions. Instead, $k_0$ is treated as a model hyperparameter, allowing the effective convective depth to be optimized during training. We also do not adopt a hybrid framework in which upper-level tendencies are computed or corrected by the original physics-based parameterization. While such approaches can improve accuracy when NN confidence is low \cite{heuer2025beyond}, they introduce additional computational cost during online simulations. Finally, to further emphasize the physically important part of the column, we apply larger loss weights to lower model levels, so that errors near the lower troposphere—where convective transport is strongest—contribute more heavily to the training objective than errors aloft.

\begin{figure}[htbp]
    \centering
    \includegraphics[scale=0.55]{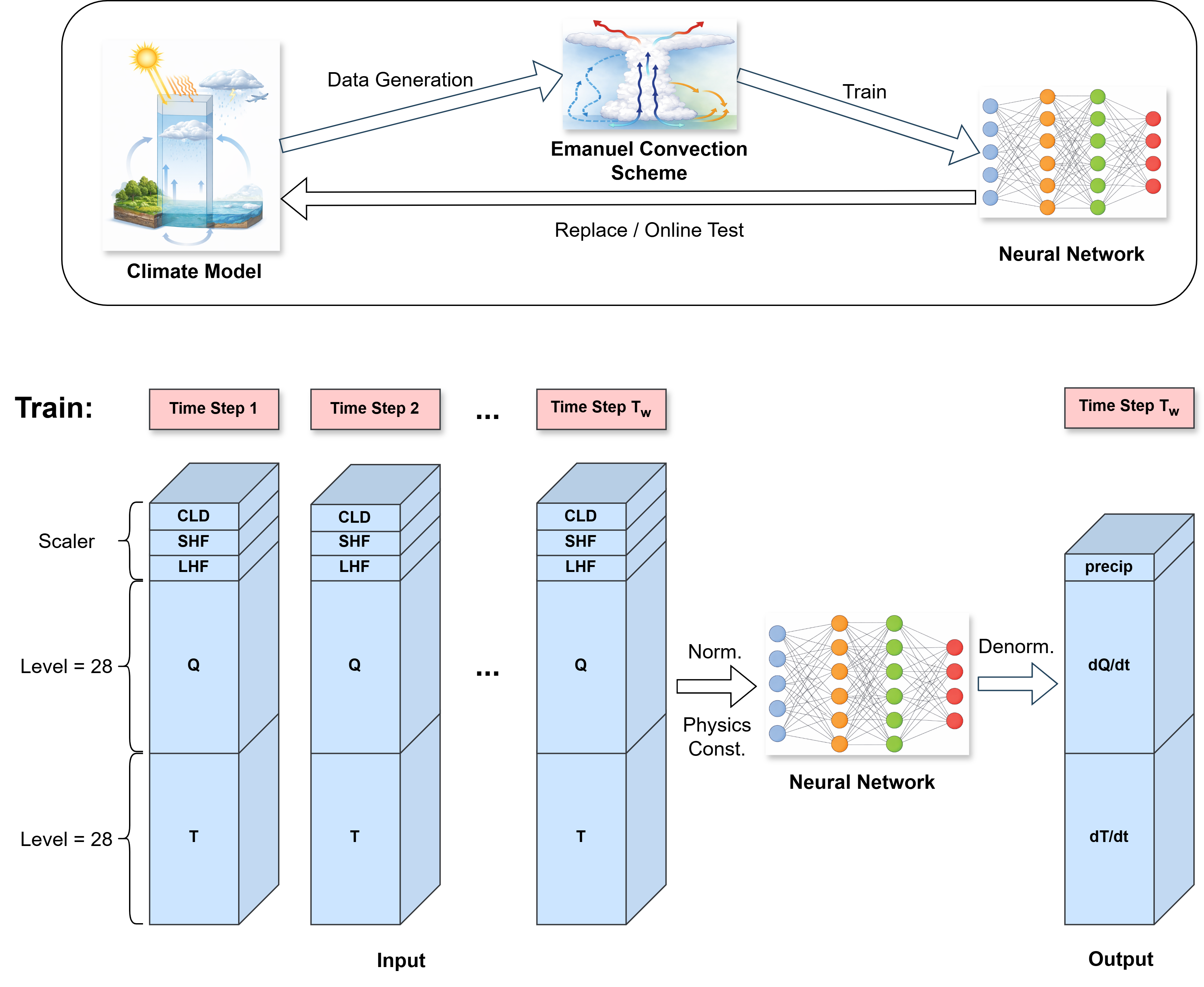}
\caption{Schematic of the NN emulation framework for the Emanuel convection scheme in a climate model. The top panel illustrates the overall workflow. A climate model is first integrated with the Emanuel convection scheme to generate training data. These data are then used to train an NN surrogate of the convection scheme. After training, the NN replaces the original Emanuel parameterization in the host climate model for online testing, allowing direct evaluation of whether the learned emulator can stably and accurately reproduce the effects of moist convection during prognostic model integration. The bottom panel shows the supervised learning setup used during training. At each sample, the NN takes as input a temporal window of atmospheric states spanning $T_w$ consecutive time steps. Each input column contains vertical profiles of temperature and specific humidity, together with other scalar surface or convective-related variables. These multi-time-step inputs are normalized before being passed into the NN, and physical constraints are applied within the learning framework to improve realism and stability.}
\label{Fig:paraformer.drawio}
\end{figure}

\subsection{Hyperparameter Search}
All models are trained and evaluated using the corresponding training and validation datasets, and the validation loss is used as the objective metric for model selection. To study the impact of convective memory on the offline performance in the Transformer architecture, we conducted several independent search processes where $T_w$ is fixed and other hyperparameters are searched freely. The search space is summarized in Table \ref{tab:2} and Table \ref{tab:tw}. In this study, we use Optuna \cite{optuna_2019}, an automated hyperparameter optimization framework, to efficiently explore the space. All models are trained with a batch size of 1024 and use a ReduceLROnPlateau learning rate scheduler with a reduction factor of 0.5 and a patience of 10 epochs. To prevent overfitting, we apply early stopping with a patience of 30 epochs on a maximum of 200 training epochs. All models are optimized using the SmoothL1 loss function with $\beta=0.1$, which balances Mean Absolute Error (MAE) and Mean Squared Error (MSE). This loss function is less sensitive to outliers than MSE while retaining stable gradients near the target values, and a similar form has been adopted in previous NN–based parameterization studies \cite{hu2025stable}. For all models, we found the best $k_0$ for all models is 19, which corresponds to a pressure of 268 hPa. This pressure is comparable to some studies, for example, \citeA{song2025physically}, which is 350 hPa, and in \citeA{hu2025stable}, which penalizes water vapor above 200 hPa.

\begin{table}[]
\centering
\caption{Hyperparameter search space for all neural network models. $T_w$ denotes the number of consecutive atmospheric states used as input in temporal models. “step” indicates that values are sampled at discrete intervals within the specified range, while “log” denotes log-uniform sampling of the initial learning rate. $k_0$ represents the hard cut-off level described in the physical constraints section. The optimal hyperparameter configuration is also reported.}
\label{tab:2}
\resizebox{0.7\textwidth}{!}{%
\begin{tabular}{lll}
\hline
\textbf{Hyperparameter}      & \textbf{Space}                       & \textbf{Best}                            \\ \hline
\multicolumn{3}{c}{\textbf{MLP}}                                                             \\ \hline
Hidden Size         & {[}128, 1024{]}, step = 128 & {[}256, 1024, 1024, 768, 640, 640{]} \\
Layers               & {[}2, 3, 4, 5, 6, 7, 8{]}   & 6                               \\ \hline
\multicolumn{3}{c}{\textbf{LSTM}}                                                            \\ \hline
Hidden Size         & {[}64, 128, 256, 512{]}     & 256                             \\
Layers               & {[}2, 3, 4, 5, 6{]}         & 3                               \\
$T_w$                  & {[}5, 10, 15, 20{]}         & 10                              \\ \hline
\multicolumn{3}{c}{\textbf{Transformer}}                                                     \\ \hline
Embedding Dimension & {[}64, 128, 256, 512{]}     &  Table \ref{tab:tw}.                              \\
Encoder Layers              & {[}2, 3, 4, 5, 6{]}         & Table \ref{tab:tw}.                                \\
$T_w$                  & {[}5, 10, 15, 20{]}         &  Table \ref{tab:tw}.                               \\
Feedforward Dimension       & {[}64, 128, 256, 512{]}     & Table \ref{tab:tw}.                              \\
Attention Heads               & {[}4, 8{]}                  & Table \ref{tab:tw}.                                \\ \hline
\multicolumn{3}{c}{\textbf{Shared Hyperparameters}}                                          \\ \hline
Optimizer           & {[}Adam, AdamW, SGD{]}      & All models: AdamW                           \\
Learning Rate       & {[}1e-4, 3e-4{]}, log       & MLP: 2.9e-4, LSTM: 1.6e-4                             \\ 
$k_0$       & {[}18, 19, 20, 21{]}       & All models: 19                               \\\hline
\end{tabular}%
}
\end{table}

\section{Offline Result}
After the best candidate models are found, we first perform an offline evaluation on the test set. In this context, offline testing refers to evaluating the trained model on pre-collected data without coupling it back into the dynamical model. The inputs and corresponding ground truth outputs are taken directly from the dataset, and the model predictions are compared against the true tendencies at each time step independently. This setup allows us to assess the predictive accuracy of the model under controlled conditions, without the influence of error accumulation or feedback from the evolving system. We use a per-level normalized Root Mean Square Error (nRMSE) metric in the offline tests because the magnitude of convective tendencies can vary strongly with height. An unnormalized metric would be dominated by levels where tendencies are naturally large, while this normalized metric can compare performance more fairly across vertical levels. The formula is:

\begin{equation}
\mathrm{nRMSE}
=
\sqrt{
\frac{1}{M}
\sum_{t=1}^{M}
\left(
\frac{
y_{t,k}^{\text{true}} - y_{t,k}^{\text{pred}}
}{
{\tilde\sigma}_{k}
}
\right)^2
}
\end{equation}
where $M$ is the size of the test set and $t$ is the index over samples (time steps). $y_{t,k}^{\text{true}}$ stands for the ground truth computed by the Emanuel convection scheme at sample $t$ and vertical level $k$, and $y_{t,k}^{\text{pred}}$ is the prediction by the NN emulator. $\tilde\sigma_k$ is the effective standard deviation of the ground truth at level $k$. We define $\tilde\sigma_k = max(\sigma_k, \epsilon)$ where $\epsilon$ is a small normalization floor to ensure numerical stability.

\begin{table}[]
\centering
\caption{The optimal hyperparameters for the Transformer with different $T_w$.}
\label{tab:tw}
\resizebox{0.7\textwidth}{!}{%
\begin{tabular}{@{}llllllll@{}}
\toprule
$T_w$ & Embedding Dim. & Encoder Layers & Feedforward Dim. & Heads & & Learning Rate & \\ \midrule
5  & 512 & 2 & 512 & 4 & & 2.3e-4 & \\
10 & 512 & 2 & 512 & 4 & & 2.1e-4 & \\
15 & 256 & 2 & 256 & 4 & & 2.9e-4 & \\
20 & 512 & 4 & 128 & 4 & & 2.2e-4 & \\ \bottomrule
\end{tabular}%
}
\end{table}

Figure \ref{Fig:unconsandcons} compares the vertical profiles of $dT/dt$ and $dQ/dt$ predicted by the unconstrained and constrained MLP models. As noted earlier, convective tendencies decrease rapidly with height and effectively vanish in the upper layers. However, the unconstrained model produces large, physically unrealistic values in these regions, where convection should be negligible. This behavior is primarily caused by artifacts introduced during normalization, which affect all levels simultaneously because the loss function is optimized over the entire column. In contrast, the constrained MLP suppresses these spurious upper-level tendencies, demonstrating that incorporating physical constraints successfully enforces the expected “no convection aloft” behavior and improves online stability.

\begin{figure}[htbp]
    \centering
    \includegraphics[scale=0.5]{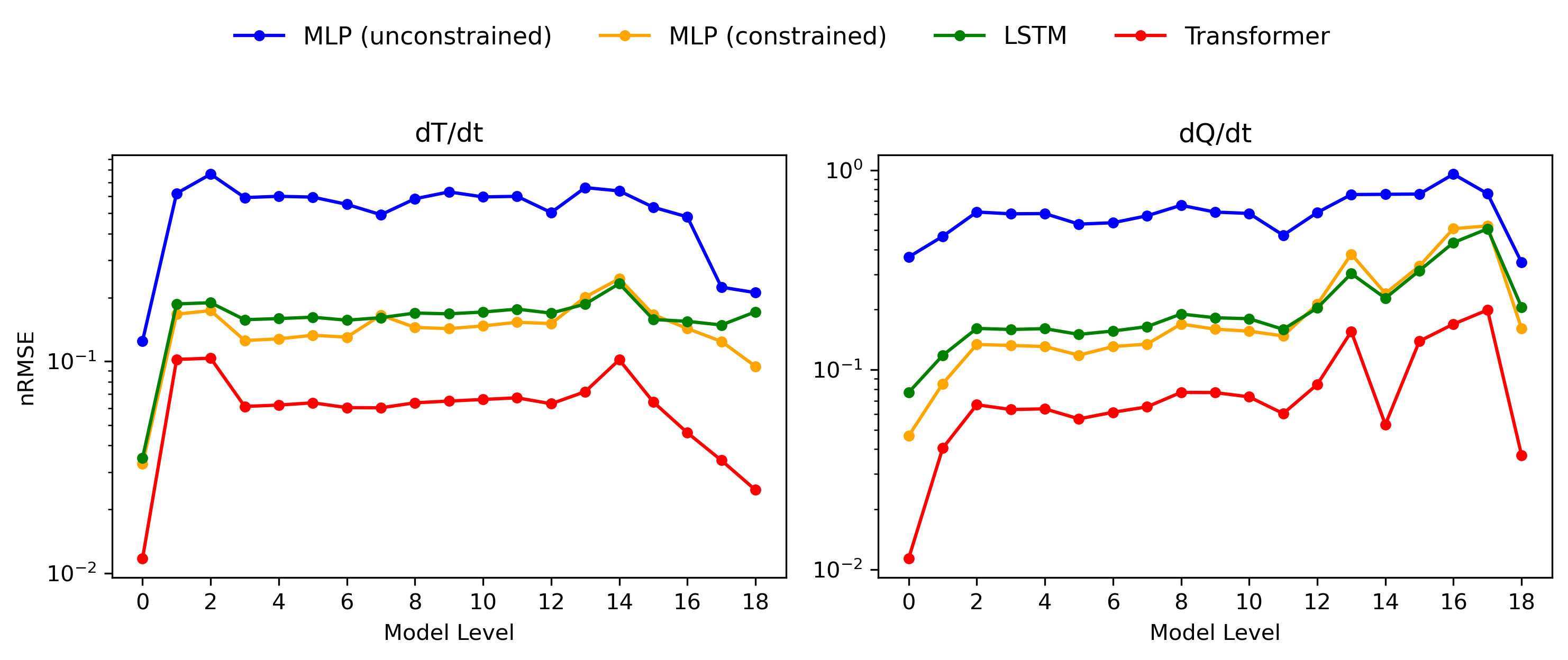}
\caption{Vertical profiles of nRMSE for heating (left) and moistening (right) tendencies in the test set. Only the lowest 19 levels are shown, as upper-level tendencies are unrealistically large in the unconstrained MLP and near zero in the constrained models. The Transformer model shown uses a temporal window of $T_w = 5$.}
\label{Fig:four_models_lev_nrmse}
\end{figure}

Figure \ref{Fig:four_models_lev_nrmse} shows the vertical distribution of nRMSE for the predicted convective tendencies. The unconstrained MLP exhibits substantially larger errors across all levels compared to the constrained models. The constrained MLP and LSTM show similar performance overall, with the LSTM slightly underperforming at mid-levels, suggesting that incorporating convective memory through a traditional sequential architecture provides only limited benefit. In contrast, the Transformer achieves the lowest nRMSE across nearly all levels for both tendencies, highlighting its superior ability to capture complex vertical and temporal dependencies. For $dT/dt$, all models show relatively lower nRMSE near the surface and in the upper levels, with larger errors at mid-levels. For $dQ/dt$, the error generally increases with height, except at the uppermost level, where all models show an abrupt drop in nRMSE. However, this sharp drop is due to the vanishing variability of moisture tendencies in the dry upper atmosphere, which caused the normalized error to become artificially small and does not reflect meaningful model skill. Figure \ref{Fig:four_models_nrmse} shows the vertically averaged nRMSE for all output variables. Consistent with Figure \ref{Fig:four_models_lev_nrmse}, the unconstrained MLP shows the largest errors, with nRMSE values of 0.55 ($dT/dt$), 0.61 ($dQ/dt$), and 0.52 ($precip$). Introducing physical constraints reduces these errors by more than 50\%, lowering nRMSE to 0.15, 0.24, and 0.11, respectively. The Transformer further improves performance, achieving reductions of approximately a factor of 5 relative to the unconstrained MLP. Among the three outputs, $dQ/dt$ remains the most challenging to predict, consistently showing the highest errors, while precipitation shows comparatively lower errors. Overall, these results demonstrate that the memory-aware Transformer provides the most accurate and robust representation of convective tendencies and associated precipitation.

\begin{figure}[htbp]
    \centering
    \includegraphics[scale=0.6]{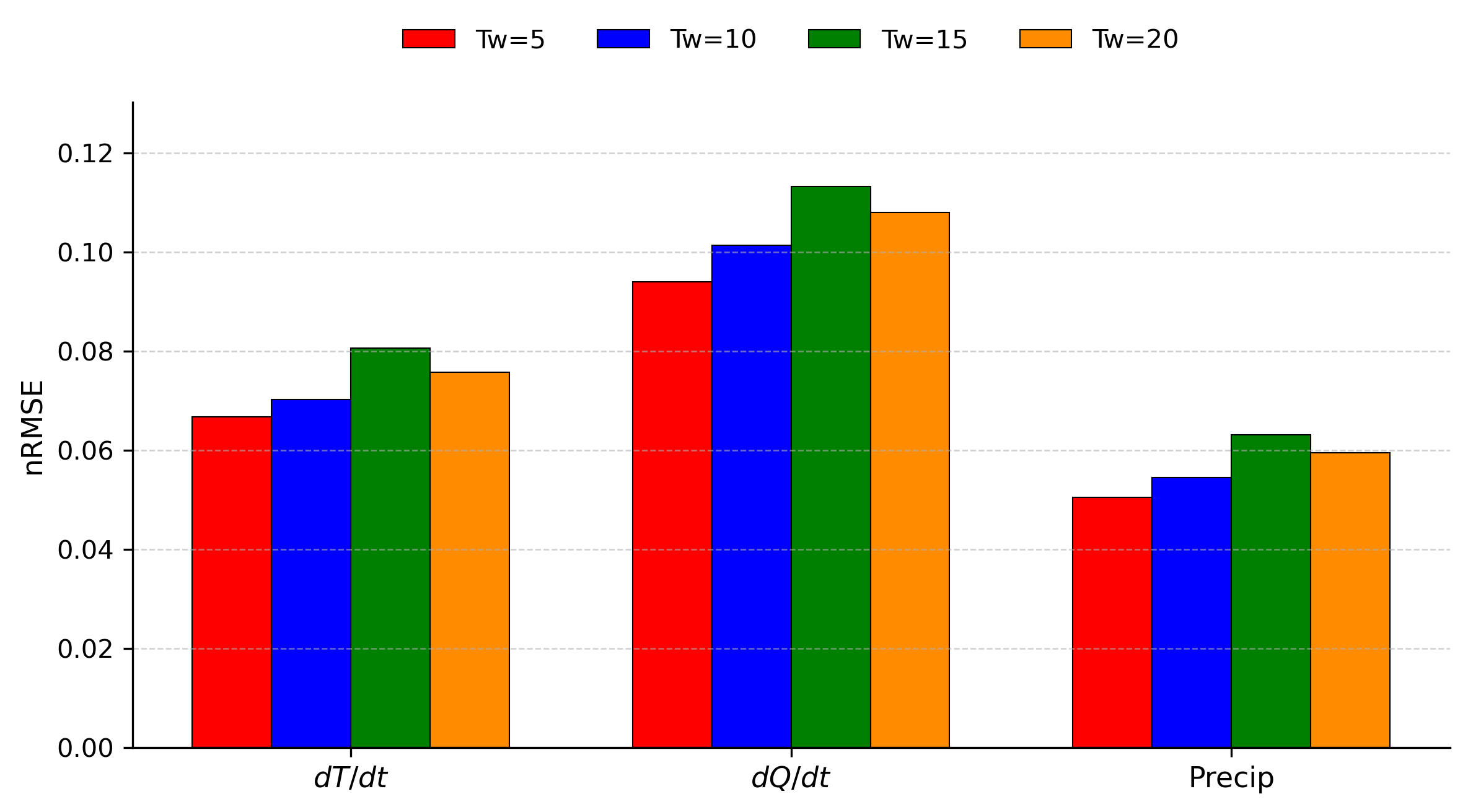}
\caption{nRMSE of temperature tendency, specific humidity tendency, and convective precipitation rate for Transformers with different temporal window lengths ($T_w$). The model with $T_w$ = 5 achieves the lowest errors across all variables, while performance degrades for longer memory ($T_w$ = 15, 20), indicating that excessive temporal context leads to error accumulation, particularly for moisture.}
\label{Fig:four_models_nrmse_trsfm}
\end{figure}

Next, we quantify the impact of convective memory length within the Transformer architecture. Figure \ref{Fig:four_models_nrmse_trsfm} presents the vertically averaged nRMSE of the outputs, showing a clear degradation in model performance as $T_w$ increases. The configuration with $T_w$ = 5, corresponding to a convective memory of approximately 50 minutes, achieves the lowest offline errors, while longer windows (up to 3 hours) lead to systematically higher prediction errors. This behavior suggests that convective processes are primarily governed by short-term dynamics, with limited dependence on extended temporal history. Figure \ref{Fig:four_models_lev_nrmse_trsfm} further illustrates the vertical structure of these errors. For $dT/dt$, the nRMSE remains relatively uniform across mid-levels but increases sharply near the upper levels (14–15), indicating greater difficulty in representing upper-tropospheric temperature tendencies. For $dQ/dt$, the error profile is more heterogeneous, with peaks around levels 13–14, highlighting challenges in modeling moisture tendencies in these regions. In conclusion, the offline results indicate that the effective convective memory in this framework is on the order of one hour, and that extending the temporal context beyond this scale degrades offline skill. This finding is consistent with \citeA{han2020moist}, which incorporates a comparable memory length of approximately 1.5 hours (5 states with a 20-minute model time step).

\section{Online Result}
Offline evaluation alone is insufficient for NN-based parameterizations because it only measures instantaneous prediction error under fixed, ground-truth inputs, without accounting for how errors accumulate and interact with the host model. In contrast, online simulation embeds the NN within the dynamical system, where its predictions feed back into future states. Even small offline errors can grow, destabilize the model, or lead to unrealistic climates. Therefore, online simulation is essential to assess stability, physical consistency, and long-term behavior, ensuring the NN parameterization performs robustly in a fully coupled simulation. In the online simulation of this study, we aim to replace the Emanuel convective scheme in climt with the trained NN emulators, which are fully coupled with other components. We conduct a 10-year online simulation from a random state in the RCE. For memory-aware models, multiple consecutive states are concatenated as inputs depending on their $T_w$.

Examples of unstable or early-terminated online simulations are shown in Figure \ref{Fig:crash}. The unconstrained MLP, due to its poor representation of tendencies near the TOA, becomes unstable almost immediately and crashes at the beginning of the simulation. The constrained MLP and LSTM eventually crashed at 2620 and 1388 time steps, respectively. The temperature and humidity evolution for the crashed models is shown in Figure \ref{Fig:crash}. The LSTM shows an early-time oscillatory instability, particularly evident in specific humidity. Large-amplitude fluctuations occur within the first 1000 steps, indicating that temporal memory introduces feedback that amplifies errors rather than damping them. This leads to a rapid departure from the physical trajectory and eventual breakdown, suggesting the LSTM is not well constrained and is sensitive to accumulated temporal errors. In contrast, the MLP appears stable for a much longer period, closely tracking the physics-based scheme for both temperature and humidity. However, it eventually experiences a late-time abrupt failure, characterized by a sudden divergence rather than gradual drift. This type of crash is typical of small, accumulated biases that remain hidden until the system crosses a stability threshold.

\begin{figure}[htbp]
    \centering
    \includegraphics[scale=0.55]{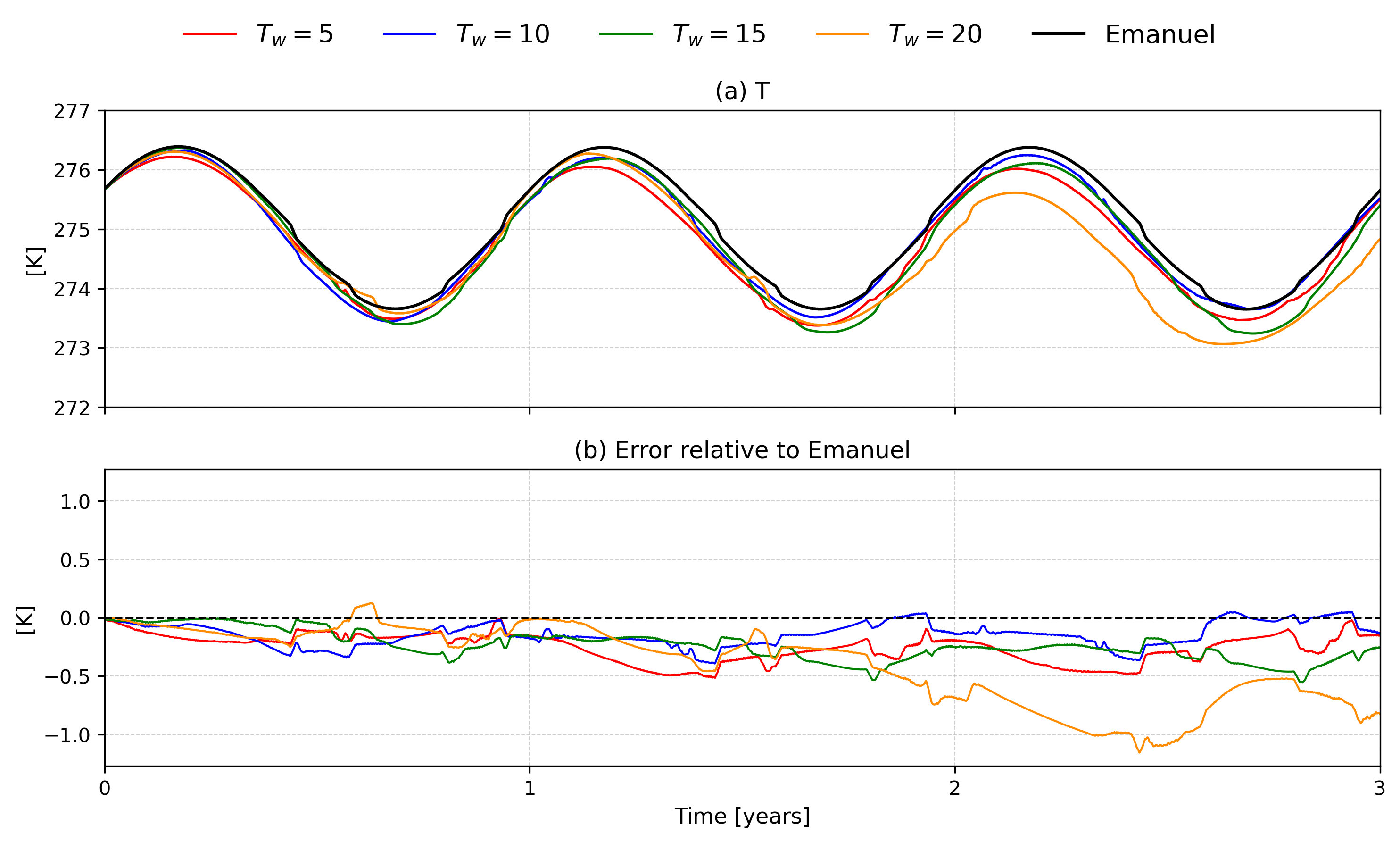}
\caption{Time series of near-surface air temperature in the first 3 years of online simulation (a) and corresponding errors relative to the Emanuel convection scheme (b) for Transformer models with different temporal memory lengths ($T_w$). A 5-day running mean is applied to reduce high-frequency variability. While all models capture the seasonal cycle, longer memory lengths exhibit larger cold biases and increased error accumulation over time.}
\label{Fig:online_t_time_3yr}
\end{figure}

\begin{figure}[htbp]
    \centering
    \includegraphics[scale=0.55]{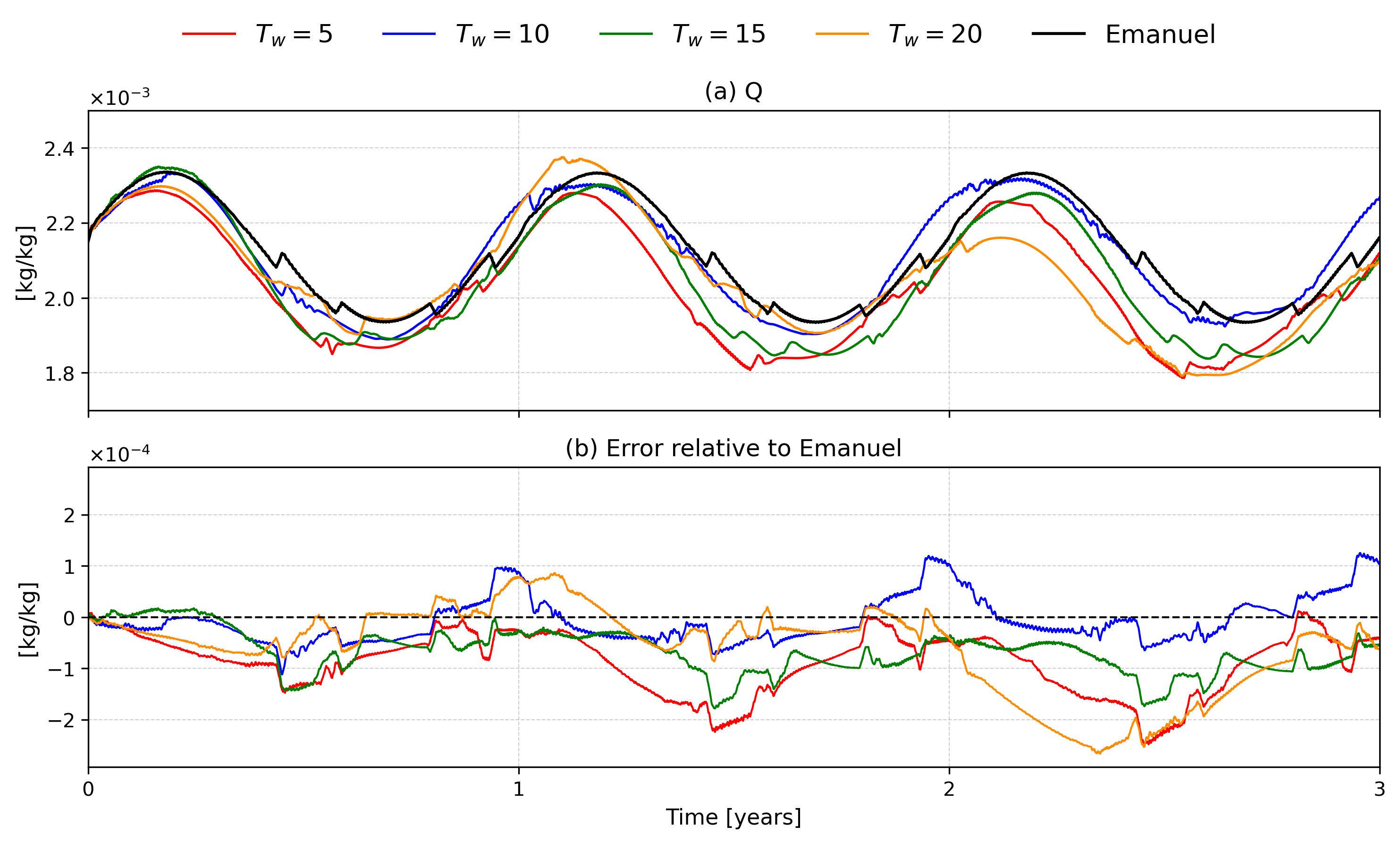}
\caption{Same as Figure \ref{Fig:online_t_time_3yr}, but for specific humidity.}
\label{Fig:online_q_time_3yr}
\end{figure}

Figure \ref{Fig:online_t_time_3yr} and Figure \ref{Fig:online_q_time_3yr} show the time evolution of surface-level temperature and specific humidity from the NN emulators compared to the physics-based reference for the first 3 years. The full 10-year time series are shown in Figure \ref{Fig:online_t_time} and Figure \ref{Fig:online_q_time}. For temperature, all models are able to capture the overall seasonal oscillation, although they generally exhibit a slight cold bias relative to the reference solution. The configurations with $T_w$ = 10 and $T_w$ = 15 track the Emanuel solution most closely, although with slightly reduced amplitudes, while $T_w$ = 5 shows larger deviations near peak temperatures. The longest-memory model ($T_w$ = 20) performs the worst overall, particularly during model years 2–4. This is consistent with its inferior offline performance, suggesting that excessive temporal memory degrades both offline accuracy and online stability. For specific humidity, which exhibits greater variability, the differences across models are more complex. The intermediate-memory models ($T_w$ = 10, 15) again provide the closest agreement with the reference, while both shorter ($T_w$ = 5) and longer ($T_w$ = 20) memory configurations underestimate peak values and display persistent positive biases, indicating an overestimation of moisture. Overall, these results suggest that an intermediate temporal window offers the best balance between capturing essential convective memory and avoiding error accumulation or over-smoothing, whereas long memory lengths lead to systematic biases in the online humidity evolution.

\begin{figure}[htbp]
    \centering
    \includegraphics[scale=0.4]{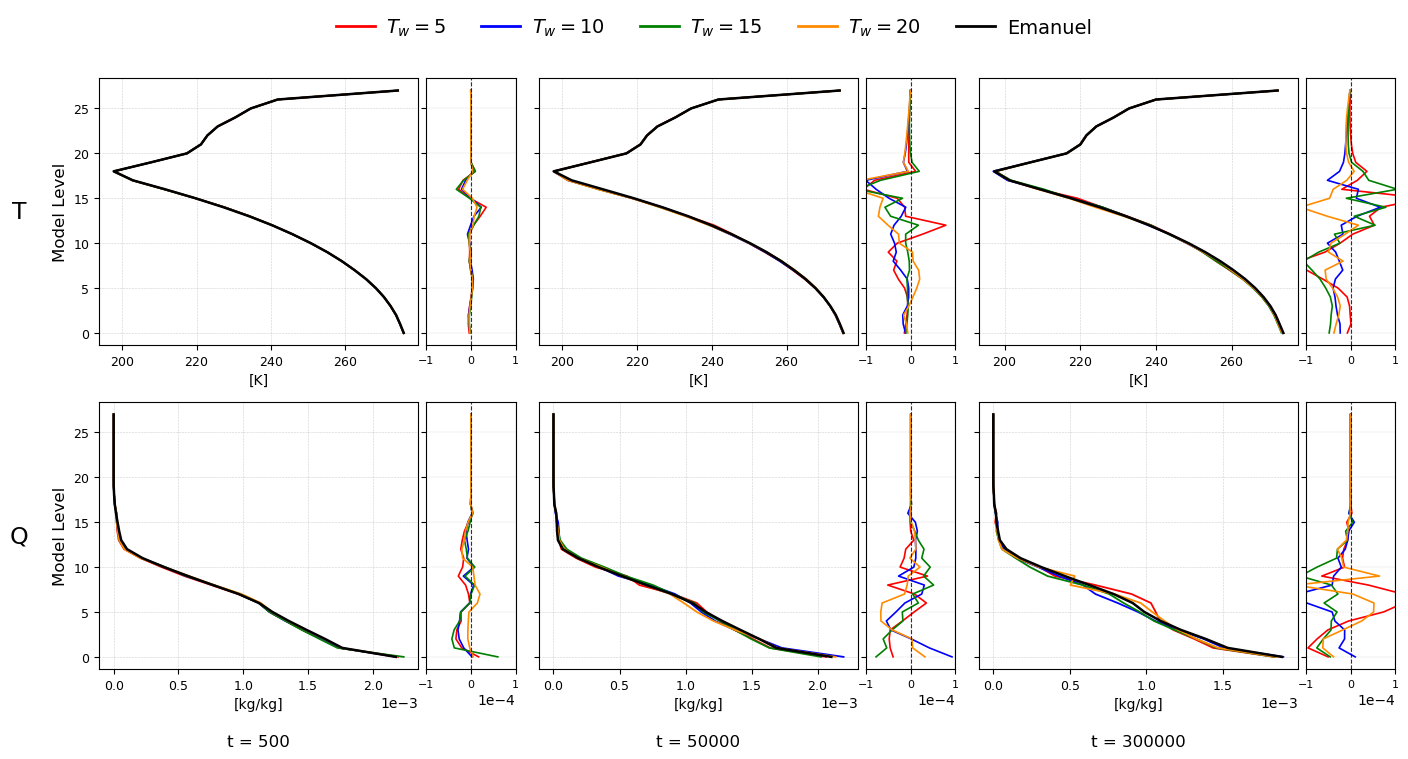}
\caption{Vertical profiles of temperature and specific humidity at three representative model time steps during the online simulation. Transformer models with different temporal memory lengths $T_w$ are compared with the Emanuel convection scheme. For each variable and time step, the main panel shows the absolute vertical profile, while the adjacent narrow panel shows the difference from Emanuel as a function of model level.}
\label{Fig:online_vert_prof_T_Q_three_times_with_diff}
\end{figure}

Figure \ref{Fig:online_vert_prof_T_Q_three_times_with_diff} and Figure \ref{Fig:online_vert_prof_dTdt_dQdt_three_times_with_diff} show the vertical profile of temperature, humidity and their tendencies in three representative time steps in the online simulation. All models reproduce the overall thermodynamic structure of the atmosphere reasonably well, but with some deviations that depend on memory length. For temperature, all Transformer configurations closely match the Emanuel reference throughout the column, with only minor differences near the surface and mid-levels, indicating strong skill in capturing the mean thermal structure. For specific humidity, small deviations appear in the mid-troposphere (around 7–10), where longer-memory models ($T_w$ = 15, 20) show slightly larger biases. In general, the vertical profile indicates that short to moderate memory ($T_w$ = 5, 10) aligns more closely with the reference tendencies, while longer memory introduces smoothing and small systematic biases, even though the mean state remains well captured.

\begin{figure}[htbp]
    \centering
    \includegraphics[scale=0.45]{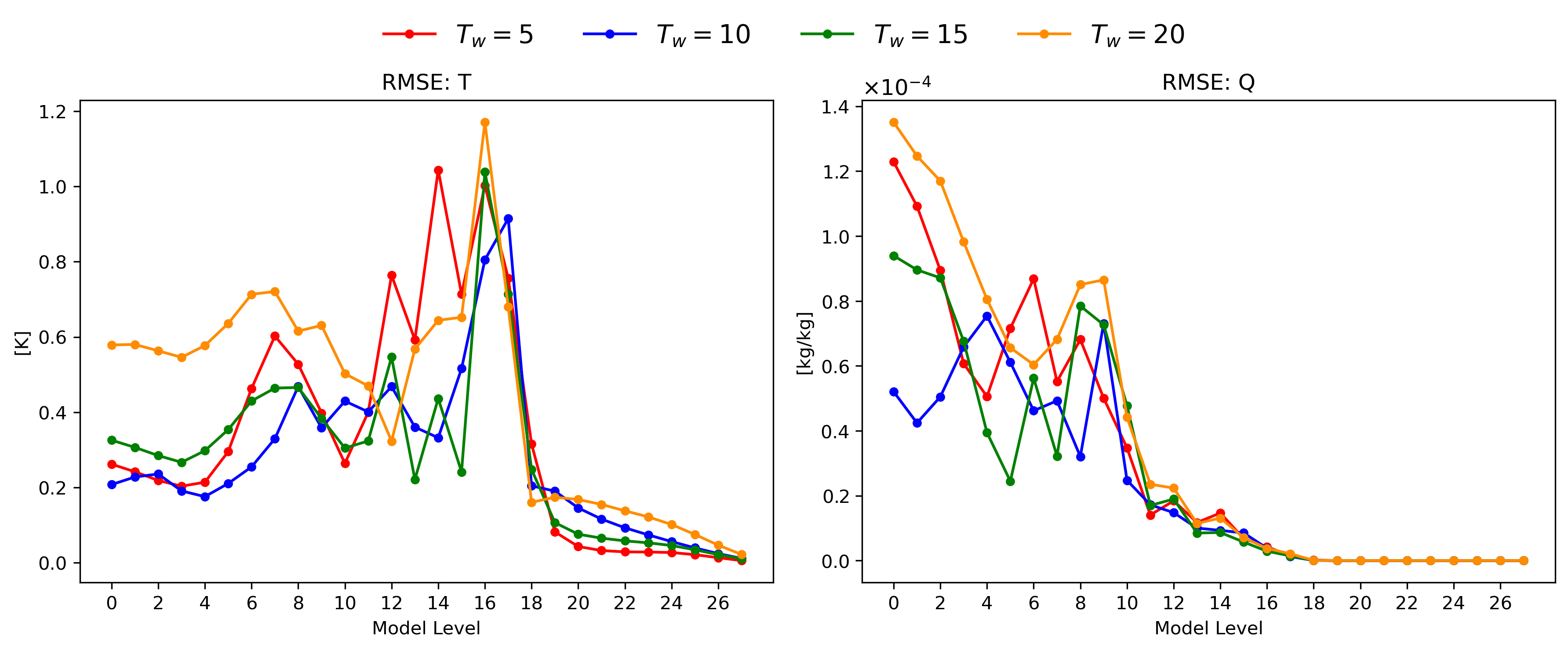}
\caption{Per-level RMSE of temperature (left) and specific humidity (right) over the 10-year online simulation for Transformer models with different temporal memory lengths ($T_w$).}
\label{Fig:online_rmse_vert}
\end{figure}

We further examine the error between the NN emulators and the Emanuel scheme on all model levels using all 10-year online simulation data, the results are shown in Figure \ref{Fig:online_rmse_vert}. The profiles of error show a strong vertical dependence, with the largest errors concentrated in the mid-to-upper troposphere (levels 12–17) for both temperature and specific humidity. For temperature, all models show relatively small errors near the surface, followed by a gradual increase and a peak around levels at 15–16, indicating difficulty in capturing convective heating and detrainment processes at these levels. Among the models, $T_w$ = 5 and 10 generally achieve lower errors, while $T_w$ = 20 consistently exhibits the highest RMSE, suggesting that longer memory degrades accuracy. For specific humidity, errors decrease more monotonically with height above level 10 and become negligible in the upper levels, but still show variability in the lower-to-mid levels where moisture processes are active. Again, shorter to moderate memory ($T_w$ = 5–15) performs better overall, while $T_w$ = 20 tends to overestimate errors across most levels. Overall, the results indicate that intermediate temporal windows provide the best performance, while excessive memory introduces systematic degradation, particularly in dynamically active layers.

\begin{figure}[htbp]
    \centering
    \includegraphics[scale=0.65]{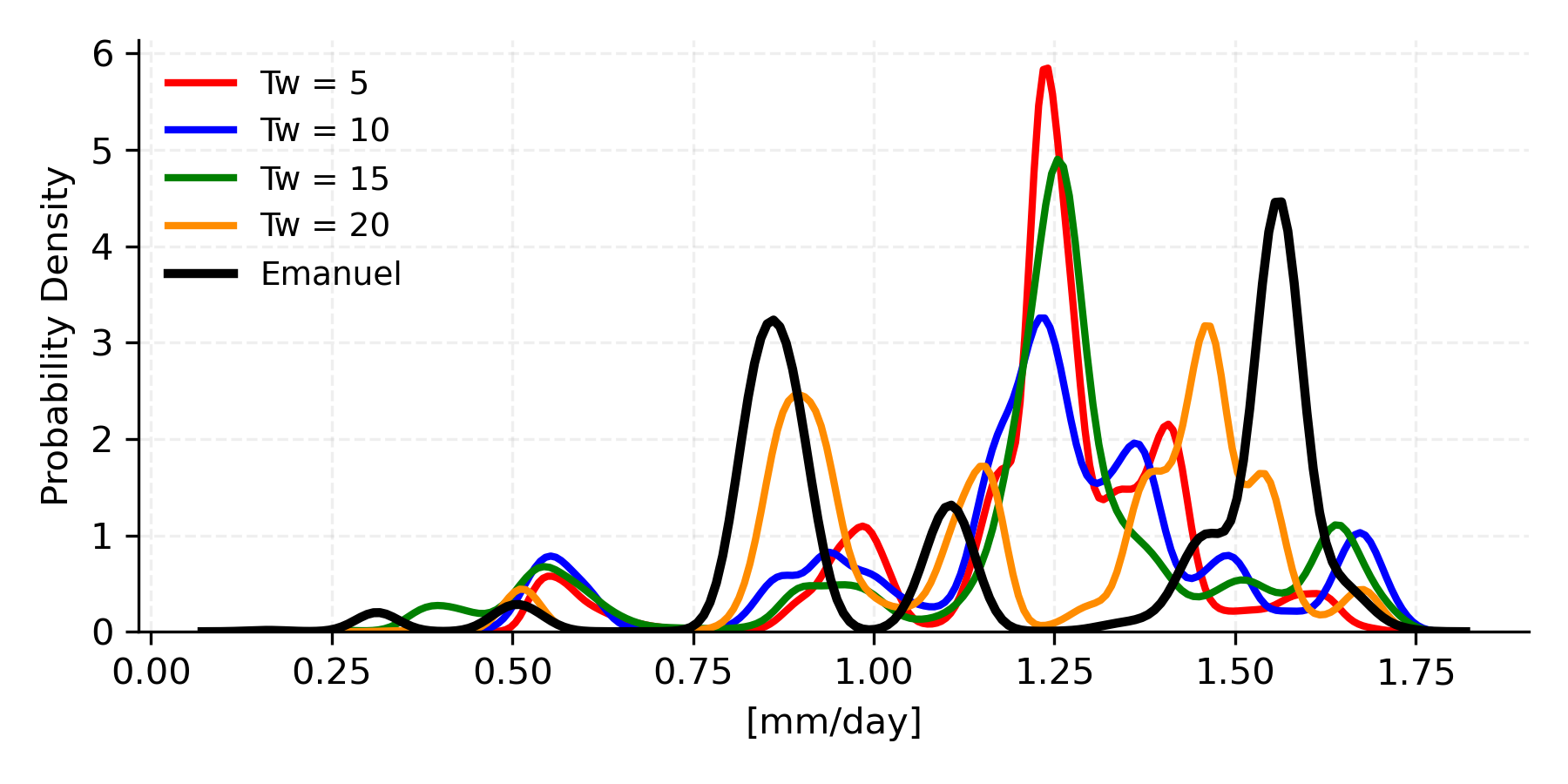}
\caption{Probability density functions of convective precipitation from the online simulation for Transformer models with different temporal memory lengths ($T_w$), compared with the Emanuel convection scheme. Differences in distribution shape indicate sensitivity of precipitation statistics to temporal memory.}
\label{Fig:online_precip_kde}
\end{figure}

Figure \ref{Fig:online_precip_kde} highlights how each model represents the intensity and variability of convective precipitation in the online simulation. The Emanuel scheme shows a multi-modal structure, with two distinct peaks around 0.8–0.9 mm/day and 1.5–1.6 mm/day, indicating separate precipitation regimes. In contrast, the NN models tend to smooth and shift these modes, with all configurations showing a dominant peak near 1.2–1.3 mm/day. The short-memory model ($T_w$ = 5) produces a sharper, more concentrated peak, while longer-memory models ($T_w$ = 15, 20) broaden the distribution and shift probability toward higher intensities, in some cases overestimating moderate-to-strong precipitation. Notably, different from the offline test, in which convective precipitation rate shows a much smaller error compared to other output variables, in the online test, none of the NN models fully capture the reference distribution, indicating a limitation in precipitation prediction. Since convective precipitation is often a thresholded, highly nonlinear diagnostic, even modest state errors can produce much larger precipitation differences online. These findings reinforce that strong offline performance does not necessarily translate to reliable online behavior \cite{yu2024climsim,ott2020fortran,wang2022stable,liu2020radnet}.

\begin{figure}[htbp]
    \centering
    \includegraphics[scale=0.4]{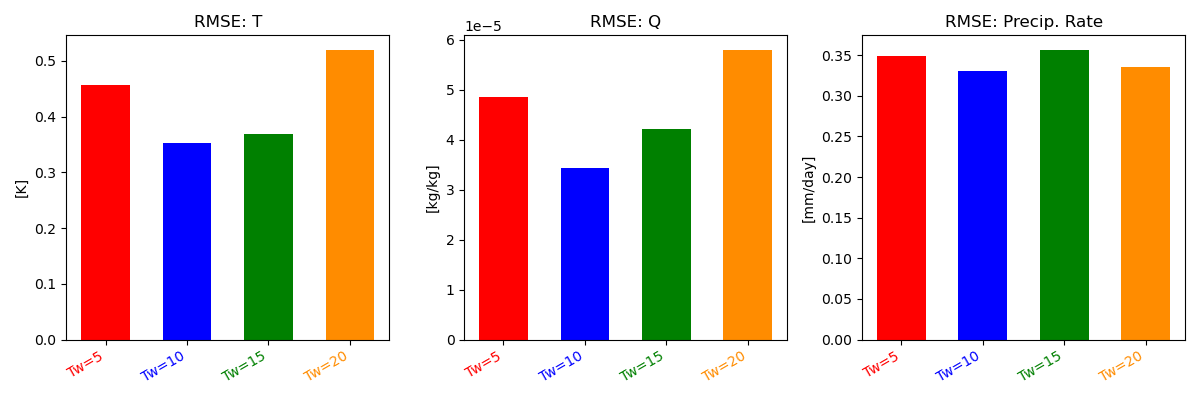}
\caption{Overall RMSE of temperature (left), specific humidity (middle), and convective precipitation rate (right) over the 10-year online simulation for Transformer models with different temporal memory lengths ($T_w$).}
\label{Fig:online_rmse_all}
\end{figure}

Figure \ref{Fig:online_rmse_all} summarizes the overall online performance of temperature, specific humidity, and convective precipitation over the 10-year simulation. For temperature, the lowest RMSE is achieved with $T_w$ = 10, indicating that a moderate temporal memory (100 minutes) provides the best balance between capturing temporal dependencies and limiting error accumulation. Both shorter ($T_w$ = 5) and longer ($T_w$ = 15, 20) memory configurations result in higher errors, with $T_w$ = 20 performing the worst, suggesting that excessive memory introduces instability or over-smoothing in the coupled system. A similar pattern is observed for specific humidity, where $T_w$ = 10 again yields the lowest RMSE, while $T_w$ = 20 produces the largest errors. This behavior highlights the sensitivity of moisture evolution to accumulated errors and indicates that longer memory windows can degrade performance, likely through amplified feedback in moist processes. For convective precipitation, the differences across models are comparatively small, with all configurations exhibiting similar RMSE values. Although $T_w$ = 10 still performs slightly better, the improvement is marginal.

\section{Discussion}
In this study, we employ an SCM to show a proof of concept that attention-based sequence models can effectively capture temporal correlations in the emulation of convective parameterization schemes. From a methodological perspective, we believe that SCMs represent an important part of the modeling hierarchy, not just for physical understanding \cite{held2005gap}, but also for the training and evaluation of NN emulators. Since SCMs cannot export heat and moisture laterally, SCM evaluations isolate errors caused by emulators. Further, since the model can only exchange heat and moisture at the surface and top-of-atmosphere, the vertical distribution of errors in Figure \ref{Fig:online_rmse_vert} shows how the model is adjusting to the errors due to the emulator: For temperature, the error maximises near the radiating height/tropopause, allowing the model to lose heat radiatively. For water vapour, the errors maximise near the surface, reducing latent heat fluxes and moisture import into the column. Thus, the impacts of the emulator's warm and wet biases on the model climate can be analysed explicitly, although a comprehensive analysis is beyond the scope of this paper.

Extending this approach to a fully coupled GCM remains an important direction for future work, particularly given the associated computational challenges. In a GCM setting with spatial grids, maintaining high temporal resolution (e.g., 10–20 minutes, same as the model time step) substantially increases training costs. To mitigate this, some prior studies have subsampled data in time \cite{yu2023climsim, mooers2021assessing}; however, such subsampling can alter the effective convective memory when temporal models are used.

Our experiments are conducted under RCE, representing a single climate regime, with both offline and online evaluations performed within this setting. As a result, the trained models are not guaranteed to generalize to different climates, model configurations, or coupled environments. Deploying NN parameterizations in broader settings will likely require training on diverse climate states to capture varying data distributions or incorporating physical constraints to ensure reasonable extrapolation during online simulations. Systematic out-of-distribution testing is therefore a critical avenue for future research. Recent work has begun exploring generalization to warmer climates \cite{han2025decadal}, but this remains an open challenge.

Another limitation arises from the quadratic computational complexity of the standard Transformer architecture used in this work, which may hinder scalability in fully coupled GCM applications. Encouragingly, a range of efficient Transformer variants has been proposed in recent years \cite{pope2023efficiently, dao2022flashattention, shen2021efficient, kitaev2020reformer}. Integrating these approaches into parameterization frameworks offers a promising path to reducing computational cost while retaining the benefits of attention-based modeling.

\section{Conclusion}
In this study, we developed a temporal memory-aware Transformer emulator for the Emanuel convective parameterization and evaluated it within an SCM using the climt framework. The results highlight the importance of physical constraints, showing that upper-level atmospheric states can degrade model performance if not properly restricted. Compared to a memory-less MLP and recurrent LSTM models, the proposed Transformer captures the nonlinear and temporally dependent behavior of moist SGS processes more effectively. Sensitivity experiments on memory length reveal that convective memory plays a critical role in both offline and online performance. An optimal memory of 100 minutes is identified, while longer temporal windows lead to degraded accuracy, likely due to accumulated errors and over-smoothing. In extended 10-year online simulations, we observe clear discrepancies between offline and online performance, reinforcing that strong offline skill does not necessarily translate to stable and accurate online behavior. Among the predicted variables, the moistening tendency remains the most challenging, consistently exhibiting larger errors than the heating tendency. Convective precipitation shows the greatest variability, with good agreement in offline tests but substantial degradation in online simulations, highlighting its sensitivity to nonlinear feedbacks and regime transitions. Overall, this work shows both the potential and the limitations of Transformer-based emulators, emphasizing the need for careful treatment of memory, physical constraints, and online evaluation in developing robust ML parameterizations.
\appendix\
\section{Supplementary Figures}
\begin{figure}[htbp]
    \centering
    \includegraphics[scale=0.5]{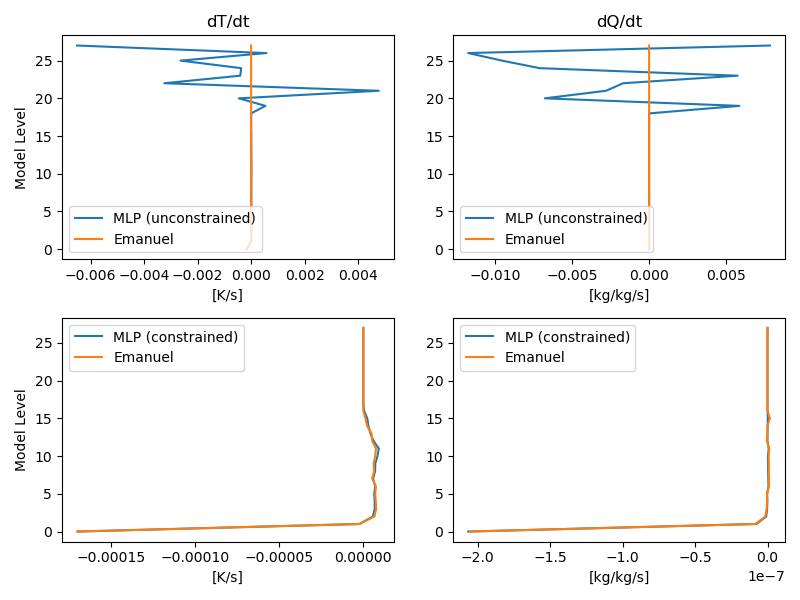}
\caption{Heating (left) and moistening (right) tendencies at one time step in the test set for unconstrained (top) and constrained (bottom) MLP, with a comparison to the Emanuel convection scheme as ground truth.}
\label{Fig:unconsandcons}
\end{figure}

\begin{figure}[htbp]
    \centering
    \includegraphics[scale=0.6]{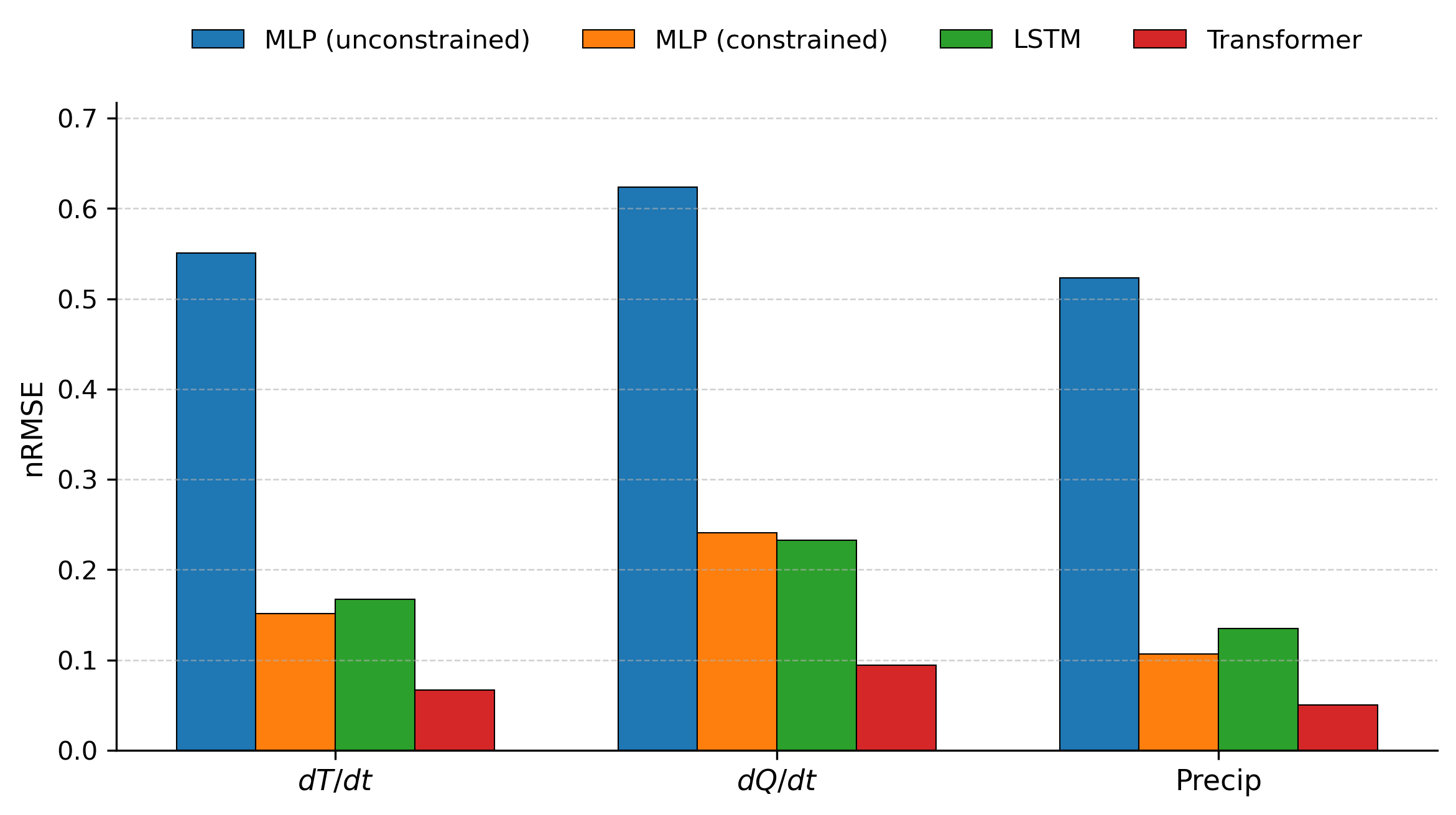}
\caption{nRMSE of temperature tendency ($dT/dt$), specific humidity tendency ($dQ/dt$), and convective precipitation rate ($precip$) for 4 NN emulators. The Transformer shown in this figure has a $T_w$ of 5.}
\label{Fig:four_models_nrmse}
\end{figure}

\begin{figure}[htbp]
    \centering
    \includegraphics[scale=0.5]{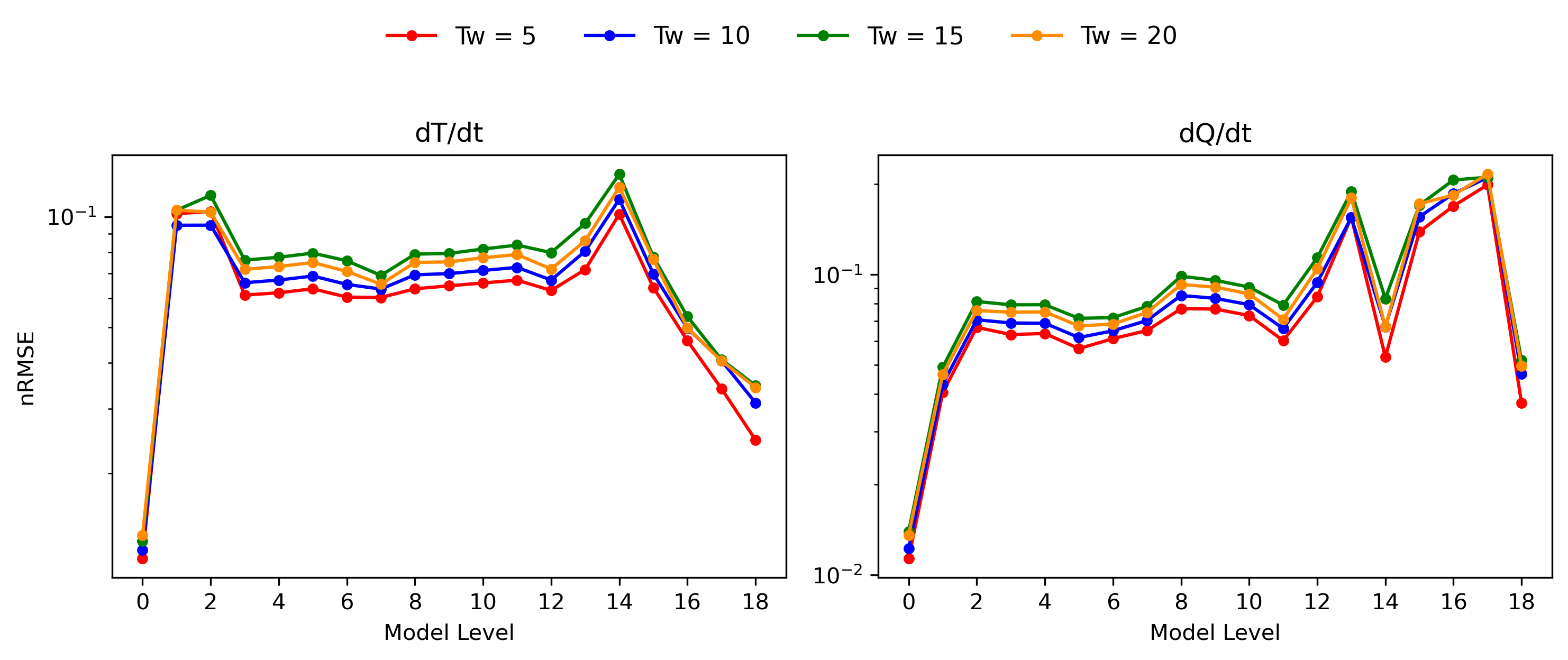}
\caption{Vertical profiles of nRMSE for heating (left) and moistening (right) tendencies in the test set for 4 Transformer models with different convective memory length ($T_w$). Only the lowest 19 levels are shown, as upper-level tendencies are unrealistically large in the unconstrained MLP and near zero in the constrained models.}
\label{Fig:four_models_lev_nrmse_trsfm}
\end{figure}

\begin{figure}[htbp]
    \centering
    \includegraphics[scale=0.4]{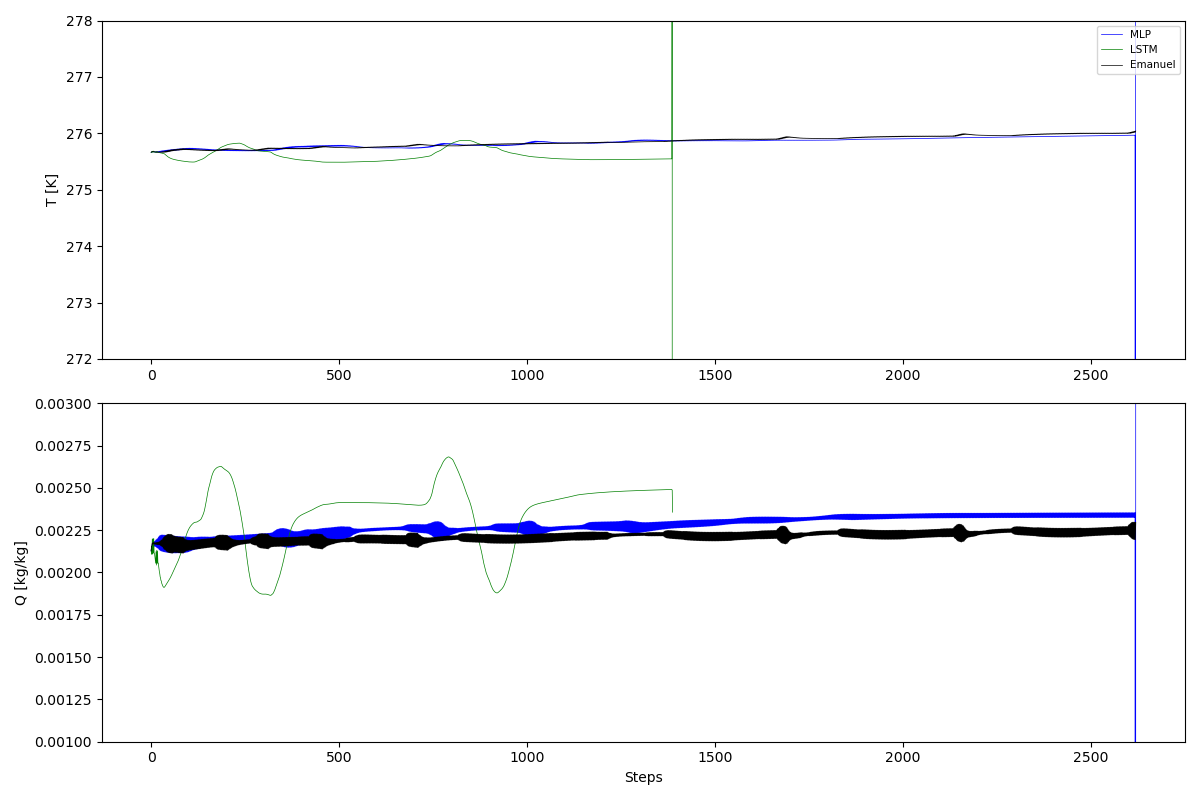}
\caption{Time evolution of temperature (top) and specific humidity (bottom) in the online simulation for MLP (blue) and LSTM (green). The LSTM exhibits early-time oscillatory instability, particularly in humidity, leading to rapid divergence and simulation crash. In contrast, the MLP remains stable and closely follows the Emanuel scheme (black) for an extended period, but eventually undergoes a sudden late-time failure.}
\label{Fig:crash}
\end{figure}

\begin{figure}[htbp]
    \centering
    \includegraphics[scale=0.5]{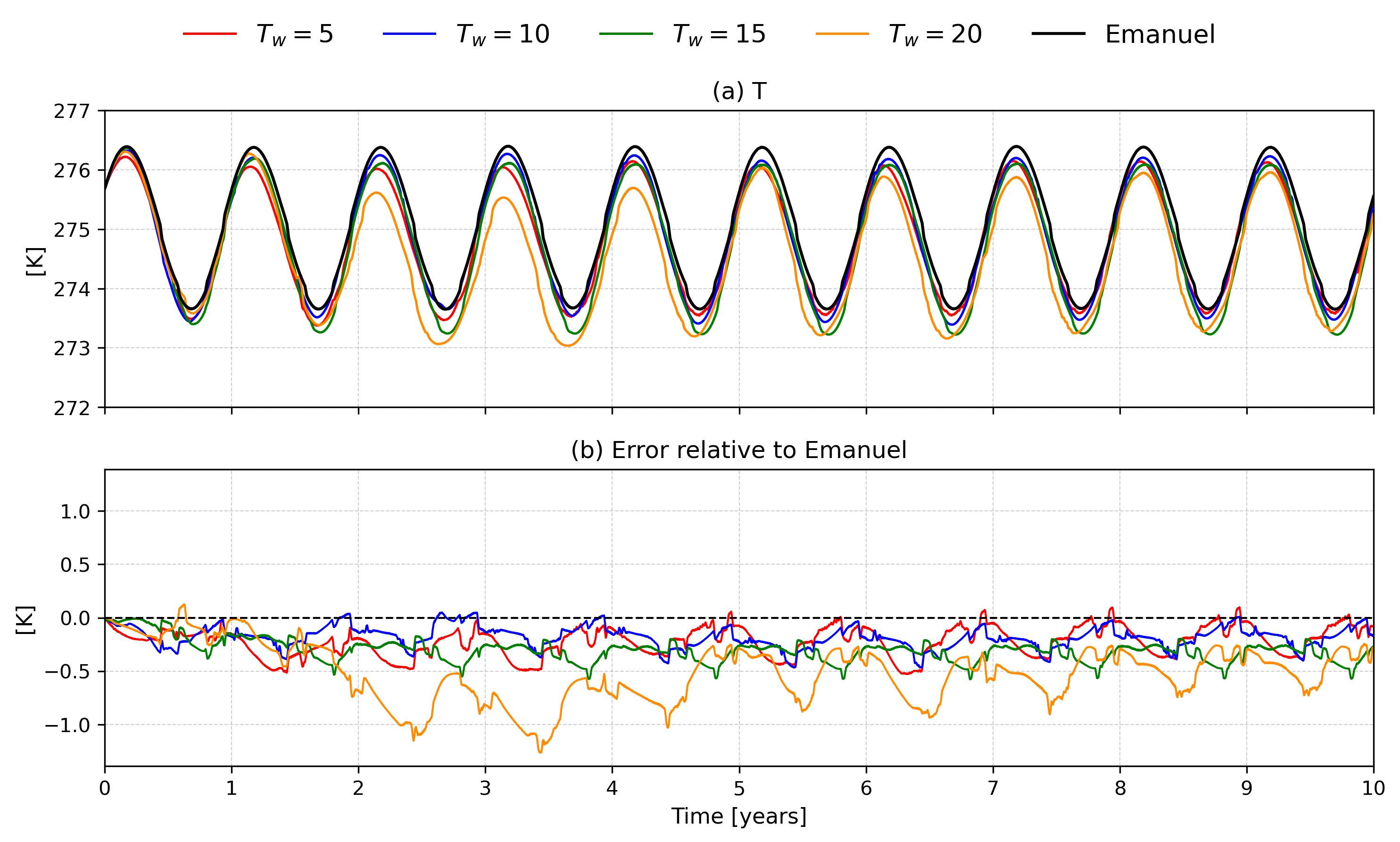}
\caption{The complete 10-year online simulation time series of surface-level temperature (a) and its error relative to Emanuel (b).}
\label{Fig:online_t_time}
\end{figure}

\begin{figure}[htbp]
    \centering
    \includegraphics[scale=0.5]{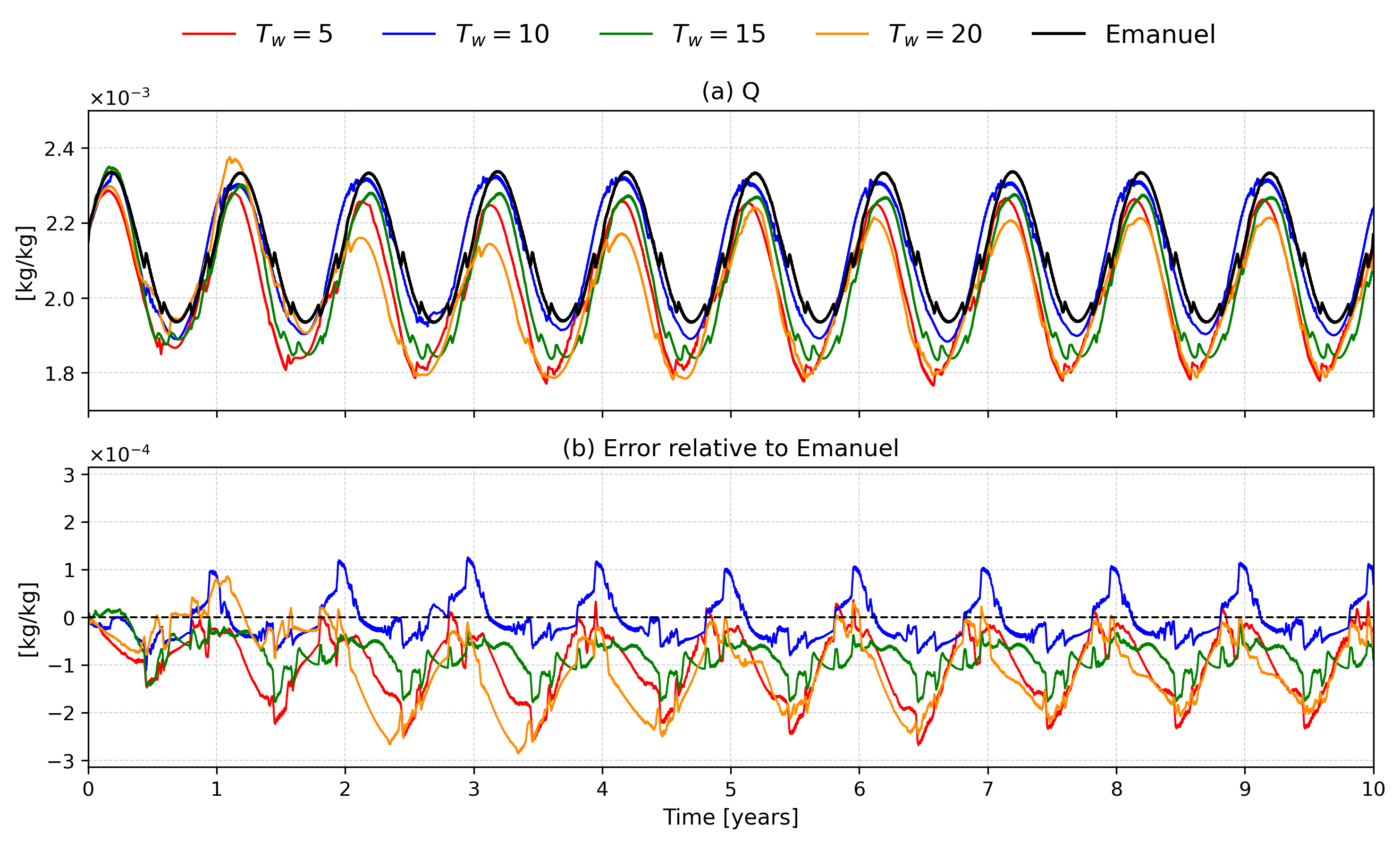}
\caption{The complete 10-year online simulation time series of surface-level specific humidity (a) and its error relative to Emanuel (b).}
\label{Fig:online_q_time}
\end{figure}

\begin{figure}[htbp]
    \centering
    \includegraphics[scale=0.4]{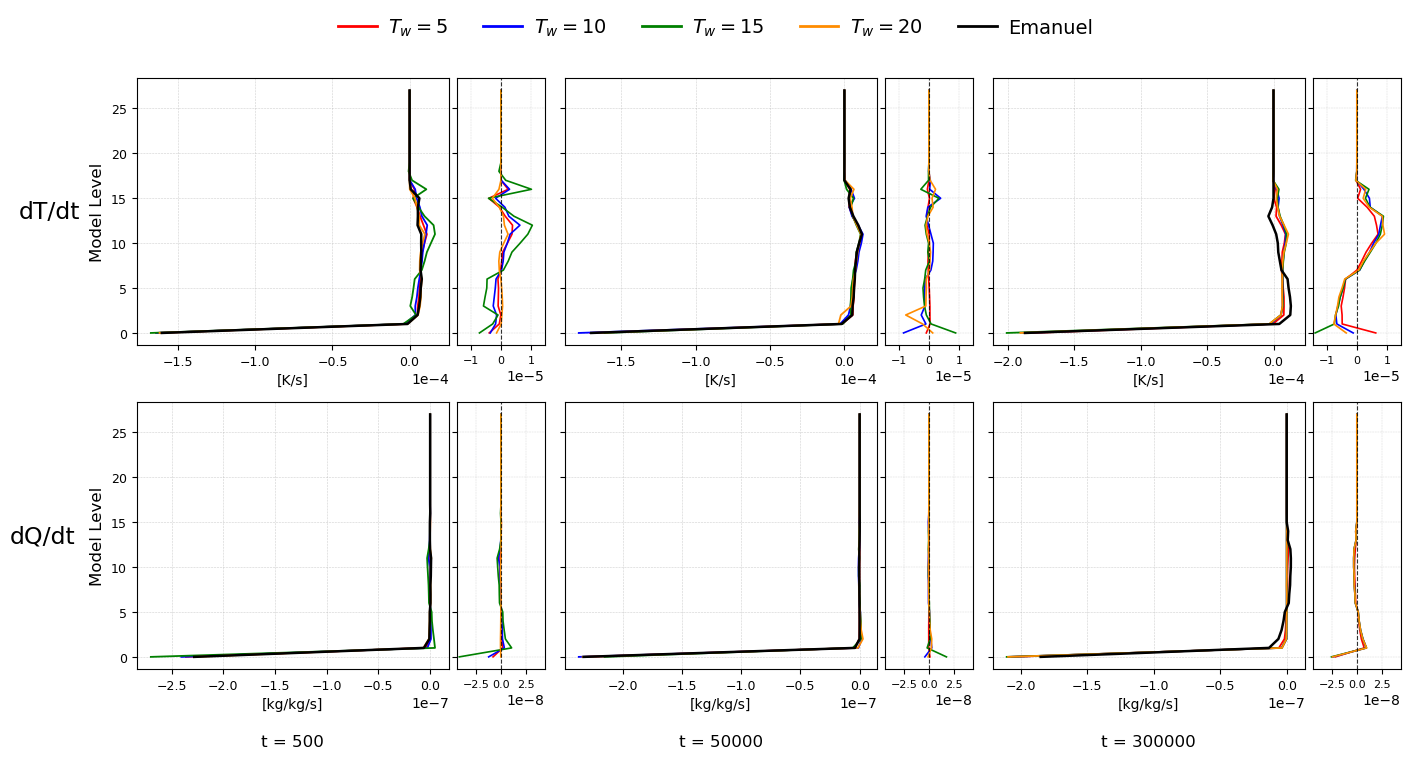}
\caption{Vertical profiles of temperature and specific humidity tendencies at three representative model time steps during the online simulation. Transformer models with different temporal memory lengths $T_w$ are compared with the Emanuel convection scheme. For each variable and time step, the main panel shows the absolute vertical profile, while the adjacent narrow panel shows the difference from Emanuel as a function of model level.}
\label{Fig:online_vert_prof_dTdt_dQdt_three_times_with_diff}
\end{figure}
%



%
%

\newpage
\section*{Open Research Section}
climt is an open-source Python library available at \url{https://github.com/CliMT/climt}. The codes used for data generation, NN training, offline and online tests, can be found at \url{https://github.com/shuochenw/climt_paraformer}.

\section*{Conflict of Interest declaration}
The authors declare there are no conflicts of interest for this manuscript.

\acknowledgments
This work was funded by the United States Department of Defense - Strategic Environmental Research and Development Program (SERDP) - RC20-1183 and the Monsoon Mission III project by the Ministry of Earth Sciences of the Government of India through the Indian Institute of Tropical Meteorology.

%
%

\bibliography{agusample}

%
%
%
%
%

\end{document}